\documentclass[reprint,aps,prl,showpacs,twocolumn]{revtex4}%
\usepackage{amsmath}
\usepackage{amsfonts}
\usepackage{amssymb}
\usepackage{graphicx}
\usepackage[colorlinks,linkcolor=blue,anchorcolor=blue,citecolor=blue,urlcolor=black]%
{hyperref}
\usepackage{mathrsfs}
\usepackage{dcolumn}
\usepackage{bm}
\usepackage{times,epsfig}%
\providecommand{\U}[1]{\protect\rule{.1in}{.1in}}

\def\be{\begin{equation}}
\def\ee{\end{equation}}
\begin{document}
\title{Optomechanically-induced-transparency cooling of massive mechanical resonators
to the quantum ground state}
\author{Yong-Chun Liu}
\author{Yun-Feng Xiao}
\email{yfxiao@pku.edu.cn}
\altaffiliation{URL: www.phy.pku.edu.cn/$\sim$yfxiao/index.html}

\affiliation{State Key Laboratory for Mesoscopic Physics and School of Physics, Peking
University; Collaborative Innovation Center of Quantum Matter, Beijing 100871,
P. R. China}
\author{Xingsheng Luan}
\author{Chee Wei Wong}
\email{cww2014@columbia.edu}
\affiliation{Optical Nanostructures Laboratory, Columbia University, New York, NY 10027, USA}
\date{\today}

\begin{abstract}
Ground state cooling of massive mechanical objects remains a difficult task
restricted by the unresolved mechanical sidebands. We propose an
optomechanically-induced-transparency cooling scheme to achieve ground state
cooling of mechanical motion without the resolved sideband condition in a pure
optomechanical system with two mechanical modes coupled to the same optical
cavity mode. We show that ground state cooling is achievable for sideband
resolution $\omega_{\mathrm{m}}/\kappa$ as low as $\sim0.003$. This provides a
new route for quantum manipulation of massive macroscopic devices and
high-precision measurements.

\end{abstract}

\maketitle





\section{Introduction}

Cavity optomechanics provides a perfect platform not only for the fundamental
study of quantum theory but also for the broad applications in quantum
information processing and high-precision metrology
\cite{RevSci08,RevRMP13,RevMeys13}. For most applications it is highly
desirable to cool the mechanical motion to the quantum ground state by
suppressing thermal noise. In the past few years numerous efforts have made
strides towards this goal through backaction cooling
\cite{GSNat11,GSNat11-2,CooNat06,CooNat06-2,CooPRL06,CooNatPhys08,CooNatPhys09-1,CooNatPhys09-2,CooNatPhys09-3,CooNat10}%
. However, the cooling limit is subjected to quantum backaction, and ground
state cooling is possible only in the resolved sideband (good-cavity) limit
\cite{PRL07-1,PRL07-2}, which requires the resonance frequency of the
mechanical motion ($\omega_{\mathrm{m}}$) to be larger than the cavity decay
rate $\kappa$. This sets a major obstacle for the ground state preparation and
quantum manipulation of macroscopic and mesoscopic mechanical resonators with
typically low mechanical resonance frequency. Therefore, it is essential to
overcome this limitation, so that ground state cooling can be achieved
irrespective of mechanical resonance frequency and cavity damping.

Some recent proposals \cite{RevCPB13} focus on circumventing the resolved
sideband restriction by using dissipative coupling mechanism
\cite{DCPRL09,DCPRL11}, parameter modulations
\cite{PulPRB09,PulPRA11-1,PulPRA11-2,PulPRL11,PulPRL12} and hybrid systems
\cite{Atom09PRA,AtomPRA13,CQEDPRL14,CoupledPRA13gxli,CoupledCLEO13ycliu,CoupledCLEO13ycliu-2}%
. Here we propose a unresolved-sideband ground-state cooling scheme in a
generic optomechanical system which does not require modified mechanisms of
coupling or specific modulation of the system parameters, or additional
components. We take advantage of the destructive quantum interference in a
cavity optomechanical system with two mechanical modes coupled to the same
optical cavity mode, where optomechanically-induced transparency (OMIT)
phenomenon \cite{OMITSCI10,OMITNat11,OMITPRA09,EIAPRA13} occurs. We show that
with the help of quantum interference, ground state cooling of the mechanical
mode with $\omega_{\mathrm{m}}\ll\kappa$ can be achieved. Moreover, we examine
the multiple input cascaded OMIT cooling which further suppresses the quantum
backaction heating. This renders quantum optomechanics with low optical-$Q$
cavities and low mechanical frequency resonators.

\section{Model}

\begin{figure}[tb]
\centerline{\includegraphics[width=\columnwidth]{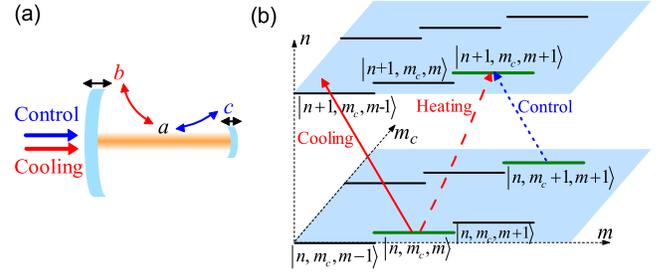}}
\caption{(color online) (a) Sketch of a typical optomechanical system with two
mechanical modes $b$ and $c$ coupled to the same optical cavity mode $a$. The
cavity is driven by a cooling laser and a control laser. (b) Energy level
diagram of the system. $\left\vert n,m_{\mathrm{c}},m\right\rangle $ denotes
the state of $n$ photons, $m_{\mathrm{c}}$ $c$-mode phonons and $m$ $b$-mode
phonons in the displaced frame. The red solid (dashed) arrow denotes the
cooling (heating) process of mode $b$. The blue dotted arrow denotes the
control laser enhanced coupling between mechanical mode $c$ and the optical
cavity mode $a$.}%
\label{Fig1}%
\end{figure}

In a generic optomechanical system, as shown in Fig. \ref{Fig1}(a), we
consider an optical cavity mode $a$ coupled to two mechanical resonance modes
$b$ and $c$, where $b$ is the mode to be cooled and $c$ is a control mode. The
cavity is driven by a cooling laser and a control laser, with frequencies
${\omega}_{0}$ and ${\omega}_{1}$, respectively. In the frame rotating at the
cavity resonance frequency ${\omega_{\mathrm{c}}}$, the system Hamiltonian
reads as
\begin{align}
H  &  =H_{b}+H_{c},\nonumber\\
H_{b}  &  ={\omega_{\mathrm{m}}b^{\dag}b+ga^{\dag}a(b+b^{\dag})+(\Omega}%
_{0}^{\ast}{{a{e}^{i\Delta_{0}t}+\mathrm{H.c.})}},\nonumber\\
H_{c}  &  ={\omega_{\mathrm{mc}}c^{\dag}c+{g}_{\mathrm{c}}{a^{\dag
}a{(c+c^{\dag})}+{{({\Omega}}}_{1}^{\ast}{ae}^{i\Delta_{1}t}}}+{\mathrm{H.c.}%
)}.
\end{align}
Here $H_{b}$ ($H_{c}$) describes the Hamiltonian related with mode $b$ ($c$);
${\omega_{\mathrm{m}}}$ (${\omega_{\mathrm{mc}}}$) is the resonance frequency
of mode $b$ ($c$); ${g}$ and ${g}_{\mathrm{c}}$ denote the single-photon
optomechanical coupling rates; ${\Omega}_{0}$ (${\Omega}_{1}$) represents the
driving strength and $\Delta_{0}={\omega}_{0}-{\omega_{\mathrm{c}}}$
($\Delta_{1}={\omega}_{1}-{\omega_{\mathrm{c}}}$) is the frequency detuning
between the cooling (control) laser and the cavity mode. For strong driving,
the linearized system Hamiltonian is given by
\begin{align}
H_{L}  &  ={\omega_{\mathrm{m}}b_{1}^{\dag}b}_{1}+{[{G}^{\left(  t\right)
}{{a_{1}^{\dag}+}}G^{\left(  t\right)  \ast}a_{1}]({b_{1}+b{{_{1}^{\dag})}}}%
}\nonumber\\
&  +{\omega_{\mathrm{mc}}{c_{1}^{\dag}}}c_{1}+[{G_{\mathrm{c}}^{{\left(
t\right)  }}{a_{1}^{\dag}+}G_{\mathrm{c}}^{{\left(  t\right)  \ast}}}%
a_{1}]({c_{1}+{{c_{1}^{\dag})}}.} \label{HL}%
\end{align}
Here the operators $a_{1}$, ${b}_{1}$ and $c_{1}$ describe the quantum
fluctuations around the corresponding classical mean fields after the
linearization; ${G}^{\left(  t\right)  }=g({{\alpha}}_{0}{e}^{-i\Delta
_{0}^{\prime}t}+{\alpha}_{1}{e}^{-i\Delta_{1}^{\prime}t})$ and ${G_{\mathrm{c}%
}^{{\left(  t\right)  }}}={{g}_{\mathrm{c}}({{\alpha}}_{0}{e}^{-i\Delta
_{0}^{\prime}t}}+{\alpha}_{1}{e}^{-i\Delta_{1}^{\prime}t})$ are the
light-enhanced optomechanical coupling strengths, with modified detunings
$\Delta_{0}^{\prime}=\Delta_{0}+\Delta_{\mathrm{om}}$, $\Delta_{1}^{\prime
}=\Delta_{1}+\Delta_{\mathrm{om}}$ and $\Delta_{\mathrm{om}}=2(g^{2}%
/{\omega_{\mathrm{m}}}+g_{\mathrm{c}}^{2}/{\omega_{\mathrm{mc}})(}\left\vert
{\alpha}_{0}\right\vert ^{2}+\left\vert {\alpha}_{1}\right\vert ^{2})$;
${{\alpha}}_{0}$ and ${{\alpha}}_{1}$ are the intracavity field {from the
contribution of the cooling and control laser inputs}; $\kappa$, $\gamma$
$(\equiv{\omega_{\mathrm{m}}/Q}_{\mathrm{m}})$ and $\gamma_{\mathrm{c}}$
$(\equiv{\omega_{\mathrm{mc}}/Q}_{\mathrm{mc}})$ are the energy decay rates of
the modes $a$, $b$ and $c$.

\section{Quantum noise spectrum}

The optical force acting on mode $b$ takes the form $F=-[{G}^{\left(
t\right)  \ast}a_{1}+{G}^{\left(  t\right)  }{{a_{1}^{\dag}]}}/x_{\mathrm{ZPF}%
}$, where $x_{\mathrm{ZPF}}$ is the zero-point fluctuation. The quantum noise
spectrum of the optical force $S_{FF}(\omega)\equiv\int dte^{i\omega
t}\left\langle F(t)F(0)\right\rangle $ is calculated to be
\begin{align}
S_{FF}({\omega}) &  =\sum\nolimits_{j=0}^{1}S_{FF}^{j}({\omega}),\nonumber\\
S_{FF}^{j}({\omega}) &  =\frac{g^{2}}{x_{\mathrm{ZPF}}^{2}}\left\vert
{{\alpha}}_{j}\tilde{\chi}_{j}\left(  {\omega}\right)  \right\vert
^{2}\nonumber\\
&  \times\lbrack\kappa+{{g}_{\mathrm{c}}^{2}}\sum\nolimits_{k=0}^{1}\left\vert
{\alpha}_{k}\right\vert ^{2}\tilde{\chi}_{\mathrm{mc}}\left(  {\omega+\Delta
}_{j}^{\prime}-{\Delta}_{k}^{\prime}\right)  ],\label{SFF}%
\end{align}
where $\tilde{\chi}_{j}^{-1}\left(  {\omega}\right)  =\chi_{j}^{-1}\left(
{\omega}\right)  +{{g}_{\mathrm{c}}^{2}}\sum\nolimits_{k=0}^{1}\left\vert
{\alpha}_{k}\right\vert ^{2}[\chi_{\mathrm{mc}}({\omega+\Delta}_{j}^{\prime
}-{\Delta}_{k}^{\prime})+\chi_{\mathrm{mc}}^{\ast}(-{\omega-\Delta}%
_{j}^{\prime}+{\Delta}_{k}^{\prime})]$, $\tilde{\chi}_{\mathrm{mc}}\left(
{\omega}\right)  =\gamma_{\mathrm{c}}(n_{\mathrm{c,th}}+1)|\chi_{\mathrm{mc}%
}({\omega)}|^{2}+\gamma_{\mathrm{c}}n_{\mathrm{c,th}}|\chi_{\mathrm{mc}%
}\left(  -{\omega}\right)  |^{2}$, $\chi_{j}^{-1}({\omega)}=-i({\omega
+{\Delta}}_{j}^{\prime}{)}+\kappa/2$ and $\chi_{\mathrm{mc}}^{-1}({\omega
)}=-i({\omega-{\omega_{\mathrm{mc}}})}+\gamma_{\mathrm{c}}/2$, with integers
$j$ and $k$ being the summation indices. Here $\chi_{j}\left(  {\omega
}\right)  $ represent the optical response to the input light and
$\chi_{\mathrm{mc}}\left(  {\omega}\right)  $ is the response function of the
control mechanical mode; $n_{\mathrm{th}}=1/[e^{\hbar{\omega_{\mathrm{m}}%
/(k}_{\mathrm{B}}T)}-1]$ and $n_{\mathrm{c,th}}=1/[e^{\hbar{\omega
_{\mathrm{mc}}/(k}_{\mathrm{B}}T)}-1]$ are the thermal phonon numbers of modes
$b$ and $c$ at the environmental temperature $T$.

\begin{figure}[tb]
\centerline{\includegraphics[width=\columnwidth]{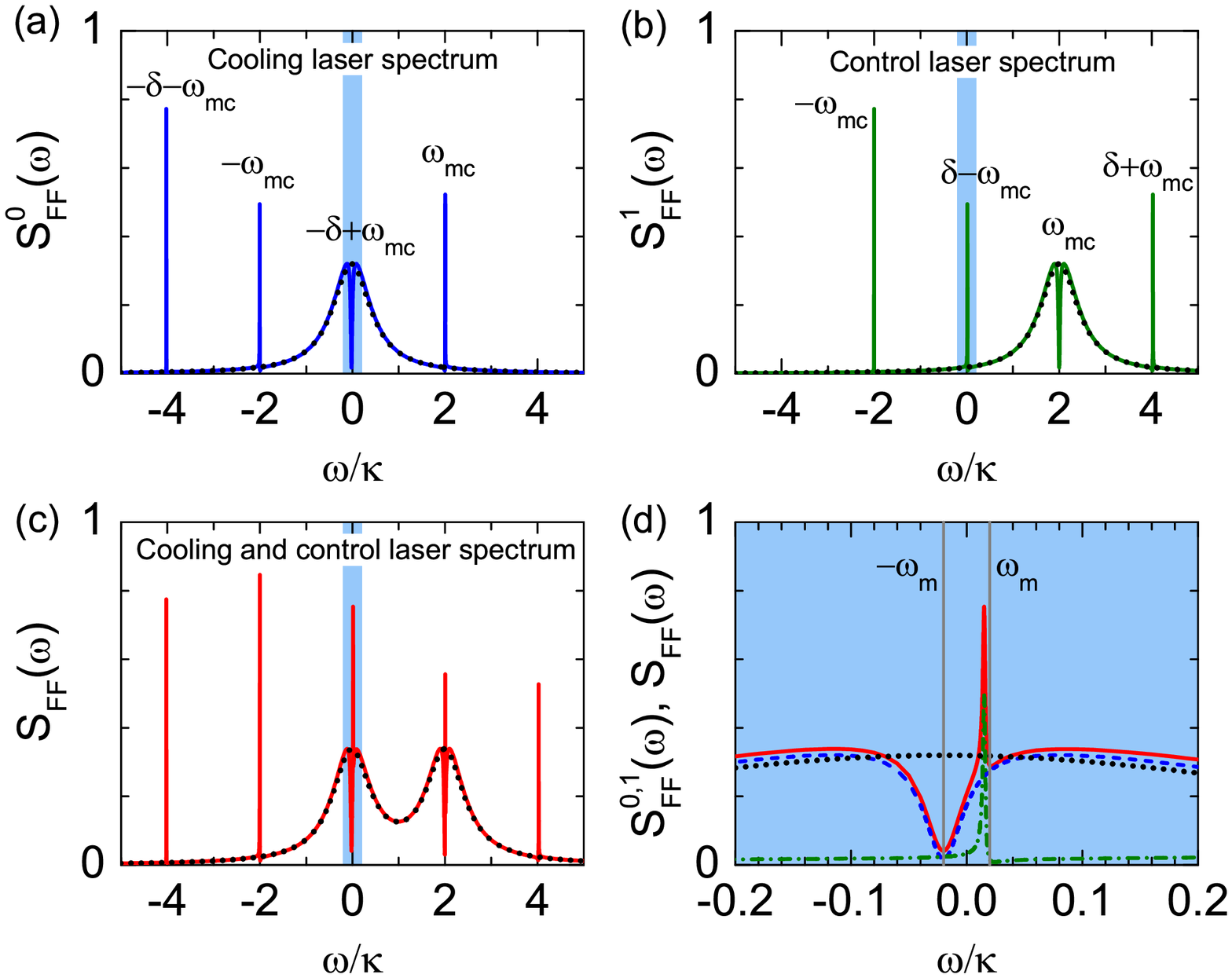}}
\caption{(color online) Quantum noise spectra (arbitrary units). The solid
curves denote (a) $S_{FF}^{0}({\omega})$, (b) $S_{FF}^{1}({\omega})$ and (c)
$S_{FF}({\omega})$ for ${\omega_{\mathrm{m}}/\kappa=0.02}$, ${\omega
_{\mathrm{mc}}/\kappa=2}$, ${\Delta}_{0}^{\prime}{=\omega_{\mathrm{m}}}$,
$\Delta_{1}^{\prime}=-{\omega_{\mathrm{mc}}}${, }$g/{\omega_{\mathrm{m}}%
=10}^{-3}$, ${{g}_{\mathrm{c}}}/{\omega_{\mathrm{mc}}=5\times10}^{-4}$,
${{\alpha}}_{0}={{\alpha}}_{1}=10^{3}$, ${Q}_{\mathrm{mc}}=10^{4}$ and
$n_{\mathrm{th}}=10^{3}$. The black dotted curves in (a)-(c) correspond to the
results without the control mode (${{g}_{\mathrm{c}}=0}$). In (a) and (b), the
position of the sharp peaks and dips are marked. (d) Zoom-in view of the
shaded region in (a)-(c). $S_{FF}^{0}({\omega})$, blue dashed curve;
$S_{FF}^{1}({\omega})$, green dashed-dotted curve; $S_{FF}({\omega})$, red
solid curve. The black dotted curves corresponds to the result without the
control mode and the control laser [$S_{FF}^{0,0}({\omega})$]. The gray
vertical lines denote $\omega=\pm{\omega_{\mathrm{m}}}$.}%
\label{Fig2}%
\end{figure}

In conventional single mechanical mode approach, the quantum noise spectrum
exhibits a standard Lorentzian curve \cite{PRL07-2}. However, here due to the
interaction between the control mechanical and the optical cavity modes, the
noise spectrum [Eq. (\ref{SFF})] is modified to a non-Lorentzian lineshape.
This originates from the quantum interference manifested by OMIT. As shown in
Fig. \ref{Fig1}(b), the system here contains a series of three-level
subsystems relevant with OMIT for heating suppression. In the presence of the
control field, the transition amplitude between the two pathways (red dashed
arrow and blue dotted arrow) destructively interfere, leading to the
suppression of the heating transition.

For the unresolved sideband regime (${\omega_{\mathrm{m}}\ll\kappa}$), the
quantum noise spectra are plotted in Fig. \ref{Fig2} with parameters
${\omega_{\mathrm{m}}/\kappa=0.02}$, ${\omega_{\mathrm{mc}}/\kappa=2}$,
${\Delta}_{0}^{\prime}={\omega_{\mathrm{m}}}$, $\Delta_{1}^{\prime}%
=-{\omega_{\mathrm{mc}}}${, }$g/{\omega_{\mathrm{m}}=10}^{-3}$, ${{g}%
_{\mathrm{c}}}/{\omega_{\mathrm{mc}}=5\times10}^{-4}$, ${\alpha}_{0}%
=\alpha_{1}=10^{3}$, ${Q}_{\mathrm{mc}}=10^{4}$ and $n_{\mathrm{th}}=10^{3}$.
Note that $S_{FF}^{0}({\omega})$ corresponds to the spectrum associated with
the cooling laser (with detuning ${\Delta}_{0}^{\prime}$) and $S_{FF}%
^{1}({\omega})$ represents the spectrum related to the control laser (with
detuning ${{{\Delta}_{1}^{\prime}}}$). Without the control mechanical mode
(${{g}_{\mathrm{c}}=0}$), they reduce to $S_{FF}^{0,0}({\omega})=\kappa
\left\vert {{\alpha}}_{0}\chi_{0}\left(  {\omega}\right)  \right\vert
^{2}g^{2}/x_{\mathrm{ZPF}}^{2}$ and $S_{FF}^{1,0}({\omega})=\kappa\left\vert
{\alpha}_{1}\chi_{1}\left(  {\omega}\right)  \right\vert ^{2}g^{2}%
/x_{\mathrm{ZPF}}^{2}$, which are Lorentzians with the centers at ${\omega
=-}\Delta_{0}^{\prime}$ and $-\Delta_{1}^{\prime}$ [black dotted curves in
Figs. \ref{Fig2}(a) and \ref{Fig2}(b)], respectively. In contrast, with
the\ presence of the control mode, a series of OMIT resonances appear in the
spectra. For $S_{FF}^{0}({\omega})$, those dips/peaks are located at
${\omega=}\pm{\omega_{\mathrm{mc}}}$, ${-\delta}\pm{\omega_{\mathrm{mc}}}$
[Fig. \ref{Fig2}(a)], where $\delta\equiv\Delta_{0}^{\prime}-\Delta
_{1}^{\prime}$ represents the two-photon detuning of the input lasers.
Thereinto, the resonances at ${\omega=}\pm{\omega_{\mathrm{mc}}}$ originate
from the interaction between the cooling laser and mode $c$, which changes the
mode density for the cavity field to absorb/emit a phonon with energy
$\hbar{\omega_{\mathrm{mc}}}$; the resonances at ${\omega=-\delta}\pm
{\omega_{\mathrm{mc}}}$ stem from the interaction among the cooling laser, the
control laser and mode $c$. Analogously, for $S_{FF}^{1}({\omega})$, the
resonances are located at ${\omega=}\pm{\omega_{\mathrm{mc}}}$, ${\delta}%
\pm{\omega_{\mathrm{mc}}}$ [Fig. \ref{Fig2}(b)]. For cooling of mode $b$, the
dips/peaks at ${\omega=}$ $\pm{\omega_{\mathrm{m}}}$ are relevant, since
$A_{\mp}\equiv S_{FF}\left(  \pm{{\omega_{\mathrm{m}}}}\right)
x_{\mathrm{ZPF}}^{2}$ are the rates for absorbing and emitting a $b$-mode
phonon by the cavity field, corresponding to the cooling and heating of mode
$b$, respectively. With the appropriate value of the two-photon detuning at
$\delta={\omega_{\mathrm{mc}}+\omega_{\mathrm{m}}}$, the OMIT lineshapes in
$S_{FF}({\omega})$ can be tuned to appear at ${\omega=}$ $\pm{\omega
_{\mathrm{m}}}$, as shown in Figs. \ref{Fig2}(c) and \ref{Fig2}(d). At
${\omega=}$ $-{\omega_{\mathrm{m}}}$, it exhibits a deep OMIT window, which
reveals the suppression of heating process, originating from the destructive
interference. Although a shallow dip also appears at ${\omega=}$
${\omega_{\mathrm{m}}}$, it only slightly decreases the mode density. The
reason is that, with ${\left\vert \Delta_{1}^{\prime}\right\vert \gg\left\vert
\Delta_{0}^{\prime}\right\vert }$, this dip is located far away from the
center (${\omega=}-\Delta_{1}^{\prime}$) of the Lorentzian background in
$S_{FF}^{1,0}({\omega})$, as shown in Fig. \ref{Fig2}(b).

\begin{figure}[tb]
\centerline{\includegraphics[width=\columnwidth]{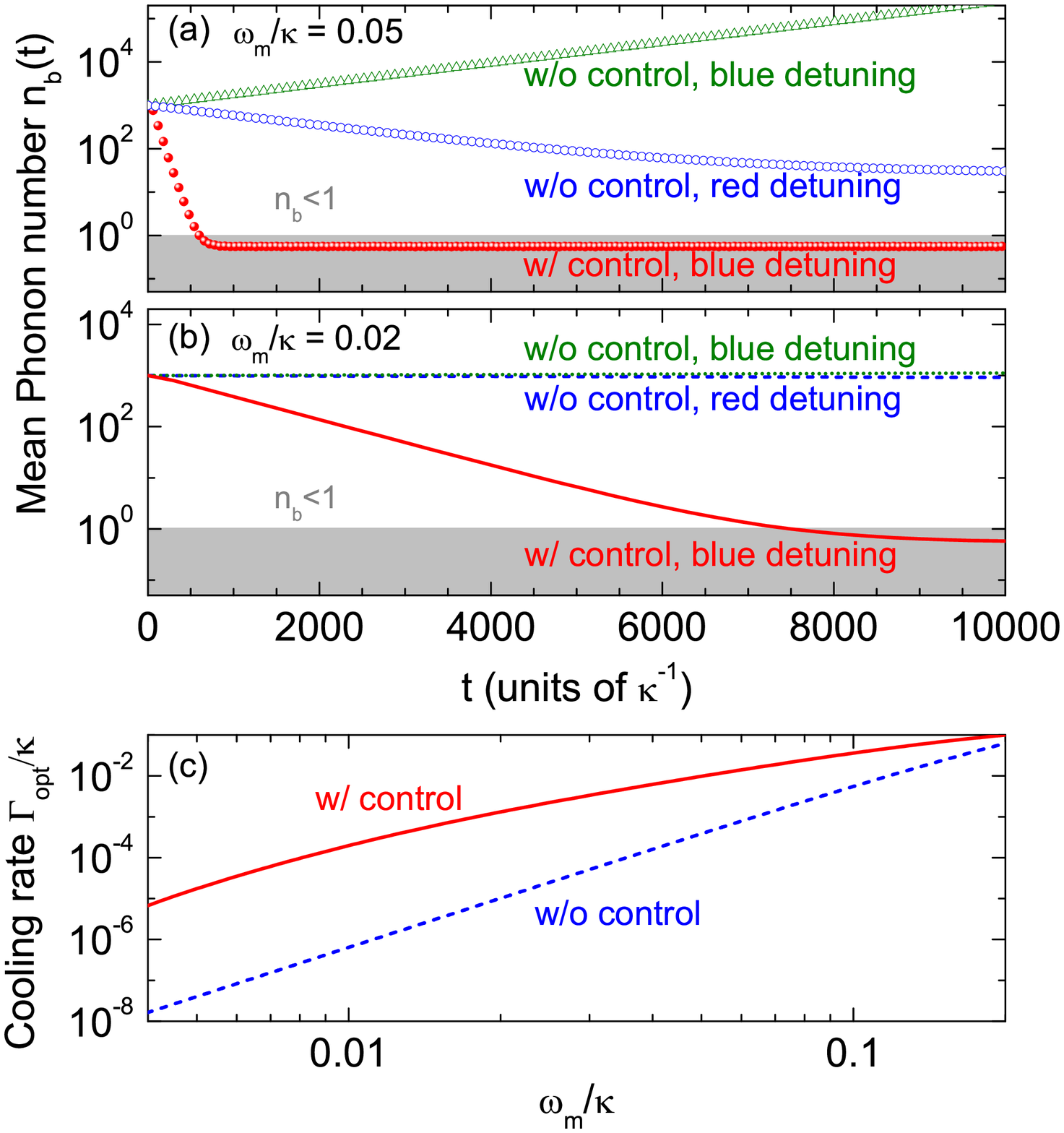}}
\caption{(color online) (a) Time evolution of the mean phonon number
$n_{b}(t)$ with control mode (red closed circle) for ${\omega_{\mathrm{m}%
}/\kappa=0.05}$, ${\omega_{\mathrm{mc}}/\kappa=2}$, ${\Delta}_{0}^{\prime
}={\omega_{\mathrm{m}}}$, $\Delta_{1}^{\prime}=-{\omega_{\mathrm{mc}}}${,
}$g/{\omega_{\mathrm{m}}=10}^{-3}$, ${{g}_{\mathrm{c}}}/{\omega_{\mathrm{mc}%
}=5\times10}^{-4}$, ${\alpha}_{0}={{1200}}$, ${{\alpha}}_{1}=1600$,
${Q}_{\mathrm{mc}}=10^{4}$, ${Q}_{\mathrm{m}}=10^{5}$ and $n_{\mathrm{th}%
}=10^{3}$. The results without the control mode and the control laser
(${{g}_{\mathrm{c}}=0}$, ${{\alpha}}_{1}=0$) for ${\Delta}_{0}^{\prime
}=-{\omega_{\mathrm{m}}}$ (blue open circle) and ${\Delta}_{0}^{\prime
}={\omega_{\mathrm{m}}}$ (green triangles) are plotted for comparison. (b)
Same as (a) except that ${\omega_{\mathrm{m}}/\kappa=0.02}$ and ${{\alpha}%
}_{0}={{\alpha}}_{1}=10^{3}$. The shaded regions in (a) and (b) denote
$n_{b}<1$. (c) Cooling rates $\Gamma_{\mathrm{opt}}$ as functions of
${\omega_{\mathrm{m}}}$ in the unit of ${\kappa}$ with the presence (red solid
curve) and absence (blue dashed curve) of the control mode and the control
laser; the parameters are the same as Fig. \ref{Fig3}(b).}%
\label{Fig3}%
\end{figure}

\section{Covariance approach}

To verify the destructive quantum interference effect, we next solve the
quantum master equation and use covariance approach
\cite{ycliuDC13,ycliuDC13-2} to obtain exact numerical results. The master
equation is given by $\dot{\rho}=i[\rho,H_{L}]+\kappa\mathcal{D}[a_{1}%
]\rho+\gamma(n_{\mathrm{th}}+1)\mathcal{D}[b_{1}]\rho+\gamma n_{\mathrm{th}%
}\mathcal{D}[{{b_{1}^{\dag}}}]\rho+\gamma_{\mathrm{c}}(n_{\mathrm{c,th}%
}+1)\mathcal{D}[c_{1}]\rho+\gamma_{\mathrm{c}}n_{\mathrm{c,th}}\mathcal{D}%
[{{c_{1}^{\dag}}}]\rho$, where $\mathcal{D}[\hat{o}]\rho=\hat{o}\rho\hat
{o}^{\dag}{{-(\hat{o}^{\dag}\hat{o}\rho+\rho\hat{o}^{\dag}\hat{o})/2}}$
denotes the the standard dissipator in Lindblad form. In Fig. \ref{Fig3}(a)
and \ref{Fig3}(b)\ we plot the time evolution of the mean phonon number
$n_{b}(t)$ for typical parameters. For single mechanical mode case in the
unresolved sideband regime, with red detuning input laser, the mechanical
motion is only slowly cooled with a small cooling rate (net optical damping
rate) $\Gamma_{\mathrm{opt}}\equiv A_{-}-A_{+}$, without reaching the ground
state. However, in the presence of the control mode and the control laser, the
cooling rate can be enhanced for more than two orders of magnitude [Fig.
\ref{Fig3}(c)], and ground state cooling with mean phonon number $n_{b}<1$ is
achievable, even for sideband resolution ${\omega_{\mathrm{m}}/\kappa}$ as
small as $0.02$. It should be emphasized that in this case the cooling laser
is blue detuned, which is quite different from the single mechanical mode
approach. For the latter, blue detuning leads to amplification instead of
cooling of the mechanical motion. The blue-detuning cooling is the unique
property originating from the quantum interference which modifies the
noise spectrum of the optical force. The OMIT lineshape can be viewed as the
inverse of the standard Lorentzian, thus the detunings for cooling are just
opposite to that for the single mechanical mode case.

\section{Cascaded OMIT cooling}

\begin{figure}[tb]
\centerline{\includegraphics[width=\columnwidth]{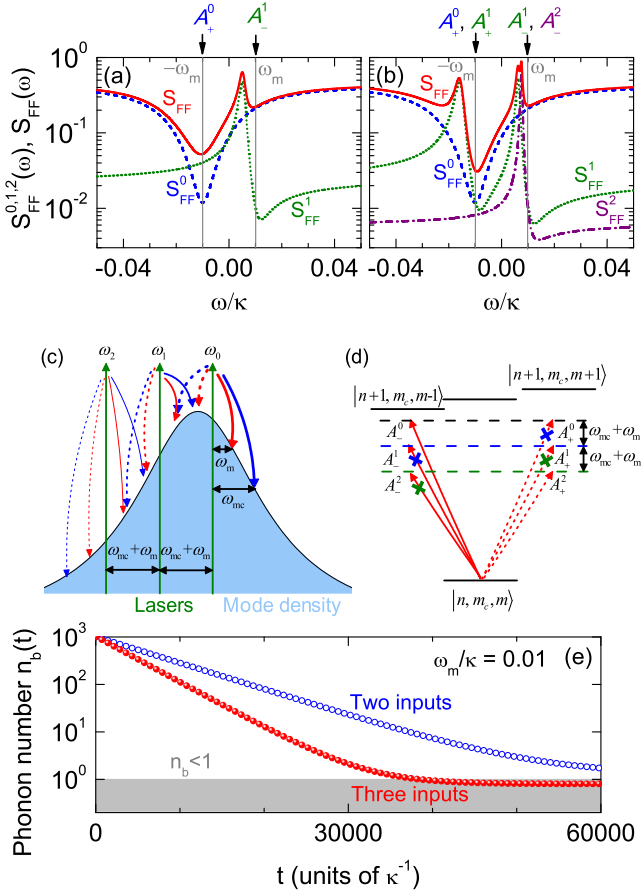}}
\caption{(color online) (a)-(b): Quantum noise spectra (arbitrary units) for
two inputs (a) and three\textbf{ }inputs (b). $S_{FF}({\omega})$, red solid
curve; $S_{FF}^{0}({\omega})$, blue dashed curve; $S_{FF}^{1}({\omega})$,
green dotted curve; $S_{FF}^{2}({\omega})$, purple dashed-dotted curve. The
gray vertical lines denote $\omega=\pm{\omega_{\mathrm{m}}}$. (c)\textbf{
}Scattering interpretation of optomechanical interactions with three inputs.
The shaded Lorentzian represents the mode density of the optical cavity. The
green vertical arrows denote the input lasers. The solid (dashed) curved
arrows denote the anti-Stokes (Stokes) scattering processes relevant with mode
$b$ (red) and mode $c$ (blue). (d) Energy levels and interpretation of the
suppression (denoted by the \textquotedblleft$\times$\textquotedblright) of
heating and cooling with three inputs. (e) Time evolution of the mean phonon
number $n_{b}(t)$ with two inputs (blue open circle) and three inputs (red
closed circle). The shaded region denotes $n_{b}<1$. Here ${{\alpha}}%
_{2}=10^{3}$, ${\omega_{\mathrm{mc}}/\kappa=2}${, }${\omega_{\mathrm{m}%
}/\kappa=0.01}$ and other parameters are the same as Fig. \ref{Fig3} (b).}%
\label{Fig4}%
\end{figure}

To further suppress quantum backaction heating, we propose the use of
additional coherent laser inputs, resulting in cascaded OMIT cooling. For $N$
inputs, the quantum noise spectrum of the optical force takes the same form as
Eq. (\ref{SFF}) except that the summation indices ($j,k$) run from $0$ to
$N-1$. As displayed in Figs. \ref{Fig4}(a)-\ref{Fig4}(d), for two inputs, the
suppression of heating for mode $b$ is the contribution of suppressed
$A_{+}^{0}\equiv S_{FF}^{0}\left(  -{{\omega_{\mathrm{m}}}}\right)
x_{\mathrm{ZPF}}^{2}$, while $A_{-}^{1}\equiv S_{FF}^{1}\left(  {{\omega
_{\mathrm{m}}}}\right)  x_{\mathrm{ZPF}}^{2}$ is also slightly suppressed. In
the presence of the third input laser with detuning $\Delta_{2}^{\prime
}=\Delta_{0}^{\prime}-2({\omega_{\mathrm{mc}}+\omega_{\mathrm{m}})}$, the
interaction involved with the control mode results in the suppression of
$A_{+}^{1}\equiv S_{FF}^{1}(-{{\omega_{\mathrm{m}}}})x_{\mathrm{ZPF}}^{2}$ and
$A_{-}^{2}\equiv S_{FF}^{2}({{\omega_{\mathrm{m}}}})x_{\mathrm{ZPF}}^{2}$.
This results in the net optical damping rate $\Gamma_{\mathrm{opt}}\equiv
\sum\nolimits_{k=0}^{2}(A_{-}^{k}-A_{+}^{k})\simeq A_{-}^{0}-A_{+}^{2}$ [Fig.
\ref{Fig4}(c) and (d)]. More generally, for $N$ inputs with detuning
$\Delta_{k}^{\prime}=\Delta_{0}^{\prime}-k({\omega_{\mathrm{mc}}%
+\omega_{\mathrm{m}})}$ ($k=1,2,...N-1$), we obtain $\Gamma_{\mathrm{opt}%
}\simeq A_{-}^{0}-A_{+}^{N-1}$. Note that the remaining heating rate
$A_{+}^{N-1}$ is much smaller than the original heating rate due to the large
detuning $\Delta_{k}^{\prime}$for the $(N-1)$-th input. In Fig. \ref{Fig4}(e)
we compare the cooling dynamics between two inputs and three inputs for
typical parameters, which shows that the cascaded OMIT cooling enables larger
cooling rate and lower cooling limit, with ground state cooling achievable
even for ${\omega_{\mathrm{m}}/}\kappa=0.01$.

\section{Cooling limits}

\begin{figure}[tb]
\centerline{\includegraphics[width=7cm]{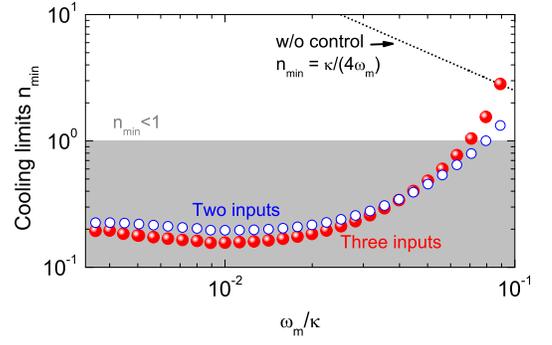}}
\caption{(color online) Fundamental cooling limits $n_{\min}$ as functions of
${{{{\omega_{\mathrm{m}}}}/\kappa}}$ for two inputs (red closed circles) and
three inputs (blue open circles). The result for single mechanical mode
approach (black dotted curve) is plotted for comparison. The shaded region
denotes $n_{\min}<1$. Here ${{\alpha}}_{0}={{\alpha}}_{1}={{\alpha}}_{2}=500$
and other parameters are the same as\ Fig. \ref{Fig4}.}%
\label{Fig5}%
\end{figure}

In Fig. \ref{Fig5} the fundamental cooling limits $n_{\min}$ as functions of
the sideband resolution ${\omega_{\mathrm{m}}/}\kappa${ }are plotted. The
exact numerical results are obtained from master equation simulations. The
black dotted curve shows the best result for conventional single mechanical
mode approach, given by $n_{\min}=\kappa/(4{{\omega_{\mathrm{m}}}})$, which is
obtained when $\Delta_{0}^{\prime}=-\kappa/2$ \cite{PRL07-1}. It reveals the
great advantage of OMIT cooling and cascaded OMIT cooling, with possibility
for ground state cooling even when ${\omega_{\mathrm{m}}/\kappa\sim3\times
10}^{-3}$, which goes beyond the resolved sideband limit by nearly $3$ orders
of magnitude. Note that Fig. \ref{Fig5} shows the cooling limits increase as
${\omega_{\mathrm{m}}/}\kappa$ increases from $\sim0.02$ to a larger value.
This is a result of blue detuning induced heating, which becomes significant
when ${\omega_{\mathrm{m}}/}\kappa$ is large.

\section{Experimental feasibility}

It should be stressed that the OMIT cooling described here adds little
complexity to the existing optomechanical system, which is crucial in the
experimental point of view. Compared with the conventional backaction cooling
approach, the additional requirement here is a control mechanical mode and one
(or more) input laser. It is experimentally feasible for various
optomechanical systems within current technical conditions. On one hand, many
optomechanical systems possess abundant mechanical modes with different
resonance frequencies, since the oscillation have different types and orders.
This situation can be found in optomechanical systems using whispering-gallery
microcavities \cite{Torid05PRL,Sphere09PRL}, photonic crystal cavities
\cite{OMcrystal09Nat,ZhengWongAPL12}, membranes \cite{Mem08Nat,BuiWongAPL12},
nanostrings \cite{Near09NatPhys} and nanorods \cite{ZhengWongOE12,LiTangNat08}
amongst others. Usually only one mechanical mode is used in most
optomechanical experiments, while exciting an additional mechanical mode is
often unintended. On the other hand, composite optomechanical systems,
containing two independent mechanical resonators, are also conceivable. For
example, in Fabry-P\'{e}rot cavities, the motion of one mirror acts as an
control mechanical mode while the other mirror is to be cooled [Fig.
\ref{Fig1}(a)]. In the near-field optomechanical system \cite{Near09NatPhys},
to cool the nanostrings, the control mode can be selected from the vibration
of the microtoroid.

\section{Conclusion}

In summary, we have presented the OMIT cooling scheme allowing ground state
cooling of mechanical resonators beyond the resolved sideband limit. It is
demonstrated that by employing the OMIT interference, quantum backaction
heating can be largely suppressed, extending the fundamental limit of
backaction cooling. The scheme is experimentally feasible, which requires
another control mechanical mode and multiple laser inputs. Such a M-O-M system
(M, mechanical mode; O, optical mode) studied here offers potential for
cooling enormous mass scale resonators \cite{LIGO09NJP,LIGO09}, which possess
small resonance frequencies. Together with the recently examined
multi-optical-mode
\cite{CoupledCLEO13ycliu,ST12NatComm,ST13SCI,ST13PRL,MM14Nat} and
multi-mechanical-mode
\cite{LinNatPhoton10,MM12NatCom,MM12PRA,MM13PRA,MM13PRA-2} systems,it is shown
that such interference effect in multi-mode cavity optomechanics\ provides
unique advantage for both fundamental studies and broad applications. Recently
we noticed a related work \cite{arXiv14}, but here we use the covariance
approach to examine the fundamental cooling limits and present detailed
analysis of cascaded OMIT cooling. This paves the way for the manipulation of
macroscopic mechanical resonators in the quantum regime.

\begin{acknowledgments}
Y.-C.L. and Y.-F.X were supported by the National Basic Research Program of
China (No. 2013CB328704, No. 2013CB921904), National Natural Science
Foundation of China (Nos. 11474011, 11222440, and 61435001), and Research Fund
for the Doctoral Program of Higher Education of China (No. 20120001110068).
X.L. and C.W.W were supported by the Optical Radiation Cooling and Heating in
Integrated Devices program of Defense Advanced Research Projects Agency
(contract number C11L10831).
\end{acknowledgments}

\end{document}